\documentclass[twocolumn]{jpsj2}

\def\runauthor{{\sc Y.~Ohta} {\it et al.}}

\title{Separated Spin and Charge Excitations and their Coupling 
in the Spin-Pseudospin Model for Quarter-Filled Ladders}

\author{
Yukinori {\sc Ohta}$^{1,2}$
\thanks{E-mail: ohta@science.s.chiba-u.ac.jp}, 
Tohru {\sc Nakaegawa}$^{2}$, and 
Satoshi {\sc Ejima}$^{2}$
\thanks{Present address: Fachbereich Physik, 
Philipps-Universit\"at Marburg, D-35032 Marburg, Germany}
}

\inst{
{$^1$Department of Physics, Chiba University, Chiba 263-8522}\\
{$^2$Graduate School of Science and Technology, Chiba University, 
Chiba 263-8522}
}

\recdate{\today}

\abst{Quantum Monte Carlo (QMC) and density-matrix 
renormalization group (DMRG) methods are used to study 
the coupled spin-pseudospin Hamiltonian in one-dimension 
(1D) that models the charge-ordering instability of the 
anisotropic Hubbard ladder at quarter filling.  
We calculate the temperature dependence of the uniform 
spin susceptibility and specific heat as well as the 
spin and charge excitation spectra of the system.  
We show that there is a parameter and temperature region 
where the spin degrees of freedom are separated from the 
charge degrees of freedom and behave like a 1D 
antiferromagnetic Heisenberg model, and that, outside 
this parameter region and above a crossover temperature, 
the spin excitations are largely affected by the charge 
fluctuation.  We argue that observed anomalous spin 
dynamics in the disorder phase of a typical charge-ordered 
material $\alpha'$-NaV$_2$O$_5$ may possibly be a 
consequence of this type of spin-charge coupling.}

\kword{charge ordering, charge fluctuation, Hubbard ladder, 
quarter filling, pseudospin}

\begin{document}
\maketitle

%%%%%%%%%%%%%%%%%%%%%%
\section{Introduction}
%%%%%%%%%%%%%%%%%%%%%%

Charge-ordering (CO) instability has recently been one 
of the major topics in the field of strongly correlated 
electron systems.  Here, elucidation of the observed 
anomalous behaviors of electrons associated with the CO 
phase transition has been the central issue.  This includes 
questions on the charge dynamics above the transition 
temperature $T_{\rm CO}$ as well as on the CO spatial 
patterns realized below $T_{\rm CO}$.  
A well-known example is the vanadate bronze 
$\alpha'$-NaV$_2$O$_5$ where the system may be modeled 
as a lattice of coupled ladders (or a trellis lattice) 
at quarter filling.\cite{smolinski,seo,nishimoto1,
mostovoy1,thalmeier}  
Strong intersite Coulomb interaction between electrons 
is believed to be the origin of the CO 
instability.\cite{seo,nishimoto1}  In this material, 
the CO with a zigzag ordering pattern is observed below 
$T_{\rm CO}=34$ K,\cite{isobe,ohama1,sawa,johnston,ohama2}  
and associated with this, a number of anomalous behaviors, 
which can be related to the slow dynamics of charge 
carriers (or charge fluctuation), have been observed 
above $T_{\rm CO}$.\cite{ravy,nakao,damascelli,presura,
ohama2,nishimoto2,nishimoto3,mostovoy2}  
Anomalous response of the spin degrees of freedom 
has also been noticed.\cite{hemberger,sushkov,johnston,
ohama3,aichhorn}
It seems therefore quite natural to wonder how in such 
systems the spin degrees of freedom behave near the CO 
phase transition when they are on the slowly fluctuating 
charge carriers.  
In this paper, we thus consider the issue: what are 
the consequences of charge fluctuation at $T>T_{\rm CO}$ 
to the spin degrees of freedom?  

One of the simplest models that allow for such situation 
is the anisotropic Hubbard ladders at quarter filling 
with the strong intersite Coulomb repulsion.   
We here use an effective Hamiltonian written in terms of 
the spin and pseudospin (representing charge degrees of 
freedom) operators.\cite{mostovoy1,cuoco,sa,mostovoy2}  
This Hamiltonian is derived from the Hubbard ladder model 
by the perturbation theory\cite{mostovoy1,cuoco,sa} where 
the hopping parameter between the rungs of the ladder is 
assumed to be small compared with the onsite and intersite 
Coulomb repulsions as well as the hopping parameter in the 
rung (i.e., the {\em anisotropic} ladder).\cite{nishimoto1}  
Although the CO is not realized in this model (since it 
is the 1D quantum-spin model), we can simulate anomalous 
behaviors of the spin degrees of freedom under the strong 
charge fluctuation.  
We will apply the quantum Monte Carlo (QMC) method to 
this model to calculate the temperature dependence of 
the uniform spin susceptibility and the spin and charge 
excitation spectra, thereby clarifying consequences of 
the interplay between its spin and charge degrees of 
freedom.  The density-matrix renormalization group (DMRG) 
method with the finite-temperature algorithm will also 
be used to calculate the temperature dependence of the 
specific heat of the model.  

We note that, in this coupled spin-pseudospin model, 
the spin exchange interaction is necessarily associated with 
the charge excitation; i.e., the spin excitations cannot 
occur without making the exchange of the pseudospins.  
We will then show that nevertheless there is a parameter 
and temperature region where the spin degrees of freedom 
behave like a 1D antiferromagnetic Heisenberg model; 
i.e., the spin degrees of freedom are `separated' from the 
charge degrees of freedom in this region.  
We will moreover show that the spin system behaves in different 
manner depending on whether the temperature $T$ is below or 
above a crossover temperature $T^*$ that is related to the 
pseudospin excitations; at $T\lesssim T^*$, it behaves like 
a 1D antiferromagnetic Heisenberg model with a $T$-independent 
effective exchange coupling constant $J_{\rm eff}$ with large 
renormalization, whereas at $T\gtrsim T^*$, $J_{\rm eff}$ 
decreases rapidly with increasing $T$, where the effective 
Heisenberg-model description ceases to be valid.  Because 
the parameter values for $\alpha'$-NaV$_2$O$_5$ are outside 
the region where the spin-charge separation is complete, 
we will argue that some experimental data may possibly be 
interpreted as consequences of this type of spin-charge 
coupling. 

This paper is organized as follows.  In \S2, we define 
the coupled spin-pseudospin model that describes the spin 
and charge degrees of freedom of the anisotropic Hubbard 
ladder at quarter filling.  Some details of the method 
of calculation are also given.  In \S3, we present the 
results of calculation which include the staggered 
susceptibility for pseudospins, the spin and pseudospin 
excitation spectra, and the temperature dependence of 
the uniform spin susceptibility and specific heat.  
Discussion on the experimental relevance to 
$\alpha'$-NaV$_2$O$_5$ and summary of the paper will 
be given in \S4.  

%%%%%%%%%%%%%%%%%%%%%%
\section{Model and Method}
%%%%%%%%%%%%%%%%%%%%%%

Our effective spin-pseudospin Hamiltonian may be written 
as a sum
\begin{equation}
{\cal H}={\cal H}_0+{\cal H}_{\rm ST}
\end{equation}
of the quantum Ising Hamiltonian for pseudospins 
\begin{equation}
{\cal H}_0=J_1\big(-\frac{g}{2}\sum_{i=1}^LT_i^x
+\sum_{i=1}^LT_i^zT_{i+1}^z\big)
\end{equation}
and the spin-pseudospin coupling term
\begin{equation}
{\cal H}_{\rm ST}=J_2\sum_{i=1}^L
\big({\bf S}_i\cdot{\bf S}_{i+1}-\frac{1}{4}\big)
\big(T_i^+T_{i+1}^-+{\rm H.c.}\big).  
\end{equation}
The standard notation is used here.  
${\bf S}_i$ and ${\bf T}_i$ are, respectively, the spin 
and pseudospin operators of spin-1/2 at site $i$, where 
$T_i^z=-1/2$ ($+1/2)$ means the electron is on the left 
(right) site on the rung of the ladder.  $L$ is the 
system size and periodic boundary condition is assumed.  
$J_1$ is the energy scale of the pseudospin system and 
$J_2$ is the coupling strength between the spin and 
pseudospin systems.  

From the second-order perturbation 
theory,\cite{mostovoy1,cuoco,sa} we have the relations 
$J_1=2V_\parallel$ and $J_2=4t_\parallel^2/V_\perp$, 
where $t_\parallel$ and $V_\parallel$ ($t_\perp$ and 
$V_\perp$) are the nearest-neighbor hopping parameter 
and Coulomb repulsion of the leg (rung) of the ladder, 
respectively.  We should then have $J_1>J_2$, which 
we assume throughout the present work.  
We also assume the onsite Coulomb repulsion to be 
$U\rightarrow\infty$.  
Relative strength of the transverse field applied to 
the pseudospins is measured by 
$g=4t_\perp/J_1=2t_\perp/V_\parallel$.  
Note that $g$ in the quantum Ising model represents 
the relative strength of the fluctuation of a charge 
in the rung: if we assume one electron in a rung, we 
have the prefactor $gJ_1/2$ in the first term of eq.~(2), 
which is the difference between the energies of the 
bonding and antibonding levels of the rung, $2t_\perp$. 
Thus, if $g$ (or $t_\perp$) is large the electron is 
stable in the bonding level of the rung, but if $g$ 
(or $t_\perp$) is small the effect of $V_\parallel$ 
easily leads the system to CO.  

We use the conventional world-line QMC method for the 
analysis of the model.  We use a 32-site cluster (where 
a site contains a spin and a pseudospin) with 
periodic boundary condition; the cluster-size dependence 
of the calculated results are examined by using clusters 
up to 96 sites but we find no significant size 
dependence in the results.  
Because the model does not conserve the total pseudospin, 
we have examined a number of ways of the spin flips 
and confirmed that available analytical results are 
reproduced correctly.\cite{nakaegawa}
The maximum-entropy method is used to calculate 
the dynamical quantities like the spin and pseudospin 
excitation spectra.  
The DMRG method with finite-temperature 
algorithm\cite{dmrg1,dmrg2} is also used for the calculation 
of the temperature dependence of the specific heat 
of our model; the method enables us to access to lower 
temperatures.  

%%%%%%%%%%%%%%%%%%%%%%
\section{Calculated Results}
%%%%%%%%%%%%%%%%%%%%%%

%%%%%%%%%%%%%%%%%%%%%%
\subsection{Staggered susceptibility for pseudospins}
%%%%%%%%%%%%%%%%%%%%%%

The response function is defined as 
\begin{equation}
\chi_{ij}=\int_0^\beta\!{\rm d}\lambda\,
\big(\langle S_j^z(-i\lambda)S_i^z\rangle
-\langle S_j^z\rangle\langle S_i^z\rangle\big)
\end{equation}
where $S_j^z(-i\lambda)$ is the Heisenberg representation 
of $S_j^z$ and $\langle\cdots\rangle$ is the canonical 
average.  $\chi_{ij}$ is Fourier transformed to the 
$q$-dependent susceptibility $\chi(q)$, which we calculate 
by the QMC method; the $q\rightarrow 0$ limit gives the 
uniform spin susceptibility $\chi(T)$ and the staggered 
susceptibility is defined as $\chi(q)$ at $q=\pi$.  
In the following, we use the subscripts ${\rm S}$ 
and ${\rm T}$ as in $\chi_{\rm S}(q)$ and 
$\chi_{\rm T}(q)$, which stand for the spin and 
pseudospin degrees of freedom, respectively.  

%%%% FIG.1 %%%%
\begin{figure}[t]
\vspace{5pt}
\begin{center}
\includegraphics[width=6.5cm,clip]{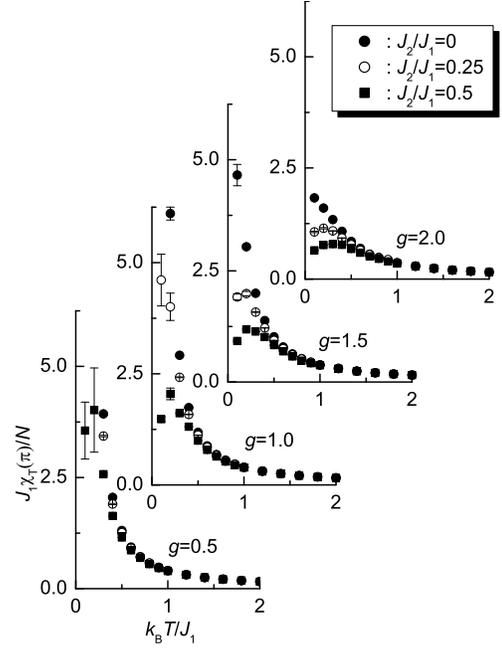}
\caption{Temperature dependence of the staggered susceptibility 
for pseudospins $\chi_{\rm T}(\pi)$ calculated for the coupled 
spin-pseudospin Hamiltonian.}
\end{center}
\label{fig:1}
\end{figure}
%%%%%%%%%%%%%%%
The phase diagram of the quantum Ising model ${\cal H}_0$ 
is well known;\cite{sachdev} at $T=0$ there is a long-range 
order for $g<1$ ($g=1$ is a quantum critical point), which 
corresponds to the zigzag (or `antiferromagnetic') CO.  
The calculated staggered susceptibility for pseudospins 
is shown in Fig.~1, where we find that it shows divergent 
behavior at $T\rightarrow 0$ for $g<1$.  
The dispersion relation of the pseudospin excitation 
observed in the calculated dynamical structure factor 
(shown in Fig.~2, see below) agrees well with the 
exact result:\cite{sachdev}  
\begin{equation}
\omega_q=\frac{J_1}{2}\sqrt{1+g^2+2g\cos q}.
\end{equation}
We find in Fig.~1 that the inclusion of the 
coupling term ${\cal H}_{\rm ST}$, which introduces the 
quantum fluctuation via the factor $T_i^+T_j^-$, suppresses 
the divergence.  Thus, we may say that the inclusion of the 
spin degrees of freedom in the quantum Ising model for 
pseudospins leads to the unstable long-range order of 
pseudospins.  

%%%%%%%%%%%%%%%%%%%%%%
\subsection{Spin and pseudospin excitation spectra}
%%%%%%%%%%%%%%%%%%%%%%

The dynamical pseudospin structure factor $S_{\rm T}(q,\omega)$ 
is defined as 
\begin{eqnarray}
&&S_{\rm T}(q,\tau)=\frac{1}{N}\sum_{ij}e^{-iq(r_j-r_i)}
\langle T_{r_i}^z(\tau)T_{r_j}^z(0)\rangle\\
&&S_{\rm T}(q,\tau)=\frac{1}{\pi}\int_0^\infty\!{\rm d}
\omega\,S_{\rm T}(q,\omega)K(\omega,\tau)\\
&&K(\omega,\tau)=e^{-\omega\tau}+e^{-\omega(\beta-\tau)}
\end{eqnarray}
where $S_{\rm T}(q,\tau)$ is the Fourier transform of the 
imaginary-time correlation function.  We use the maximum 
entropy method for the inverse Laplace transformation (or 
analytical continuation) to obtain $S_{\rm T}(q,\omega)$ 
from $S_{\rm T}(q,\tau)$.  
The dynamical spin structure factor $S_{\rm S}(q,\omega)$ is 
similarly defined by replacing the pseudospin operator $T^z_r$ 
with the spin operator $S^z_r$.  

%%%% FIG.2 %%%%
\begin{fullfigure}[t]
\vspace{5pt}
\begin{center}
\includegraphics[width=11.0cm,clip]{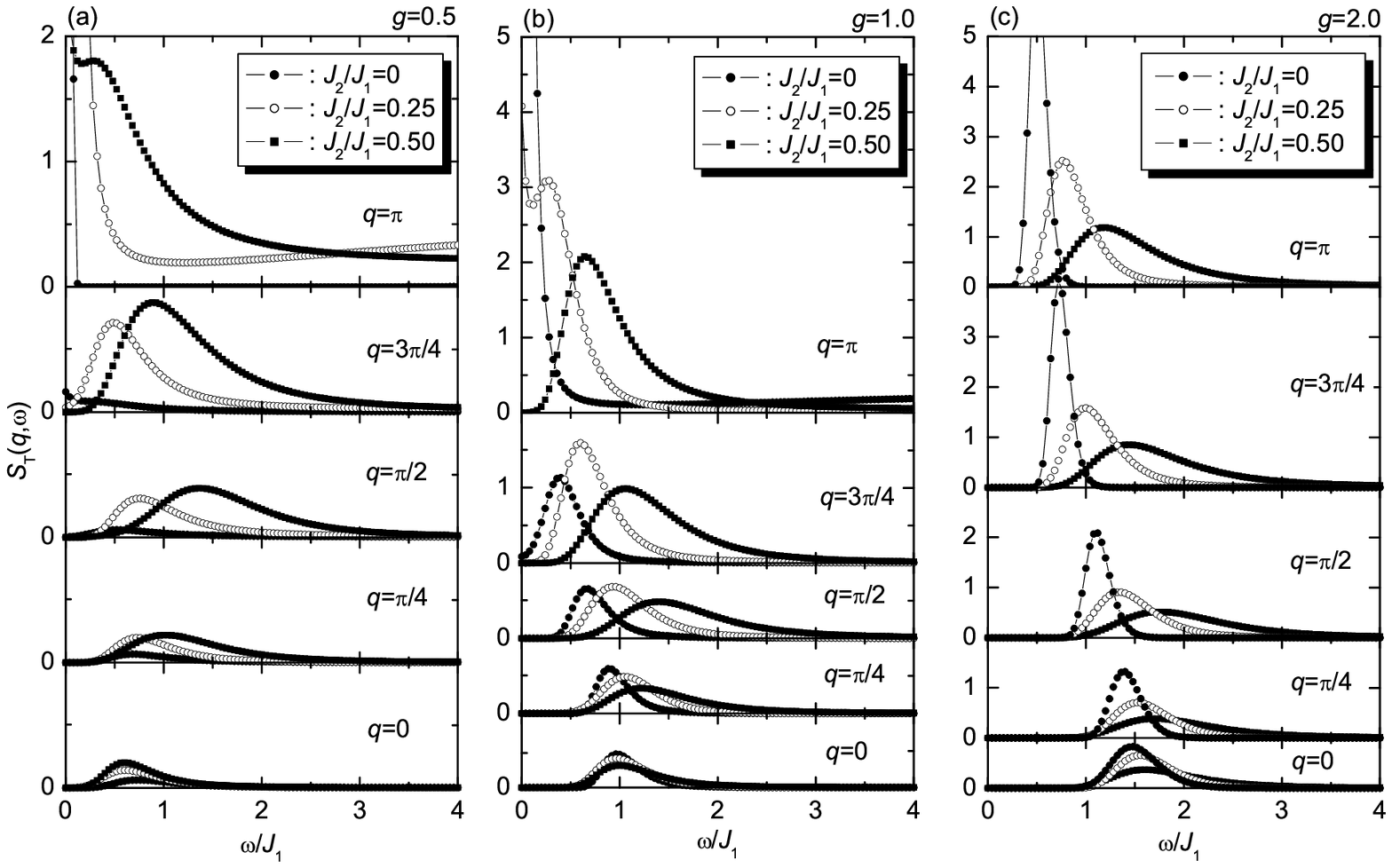}
\caption{Dynamical pseudospin structure factor 
$S_{\rm T}(q,\omega)$ for the coupled spin-pseudospin 
model calculated at $k_{\rm B}T=0.1J_2$.  
The results at $J_2/J_1=0$ are for the quantum Ising 
model.  The peak at $\omega=0$ for $J_2/J_1>0$ 
in the uppermost panel of (a) and (b) is spurious, which 
is due to the error of the maximum entropy method.}
\end{center}
\label{fig:2}
\end{fullfigure}
%%%%%%%%%%%%%%%
The calculated results for the pseudospin excitation spectra 
at low temperature ($k_{\rm B}T=0.1J_2$) are shown in Fig.~2, 
where we find that the spectra are under strong influence of 
the spin-pseudospin coupling term $J_2$.  
With increasing the coupling strength $J_2/J_1$, the peak of 
the pseudospin spectra shifts to higher energies and 
simultaneously the spectra are broadened.  
Thus, the lower-energy edge of the peak is not affected 
strongly by the coupling strength $J_2$, at least when $g$ 
is large.  It seems reasonable to suppose that the scattering 
of the pseudospin excitations due to spin excitations causes 
the broadening of the spectra.  

%%%% FIG.3 %%%%
\begin{fullfigure}[t]
\vspace{5pt}
\begin{center}
\includegraphics[width=11.0cm,clip]{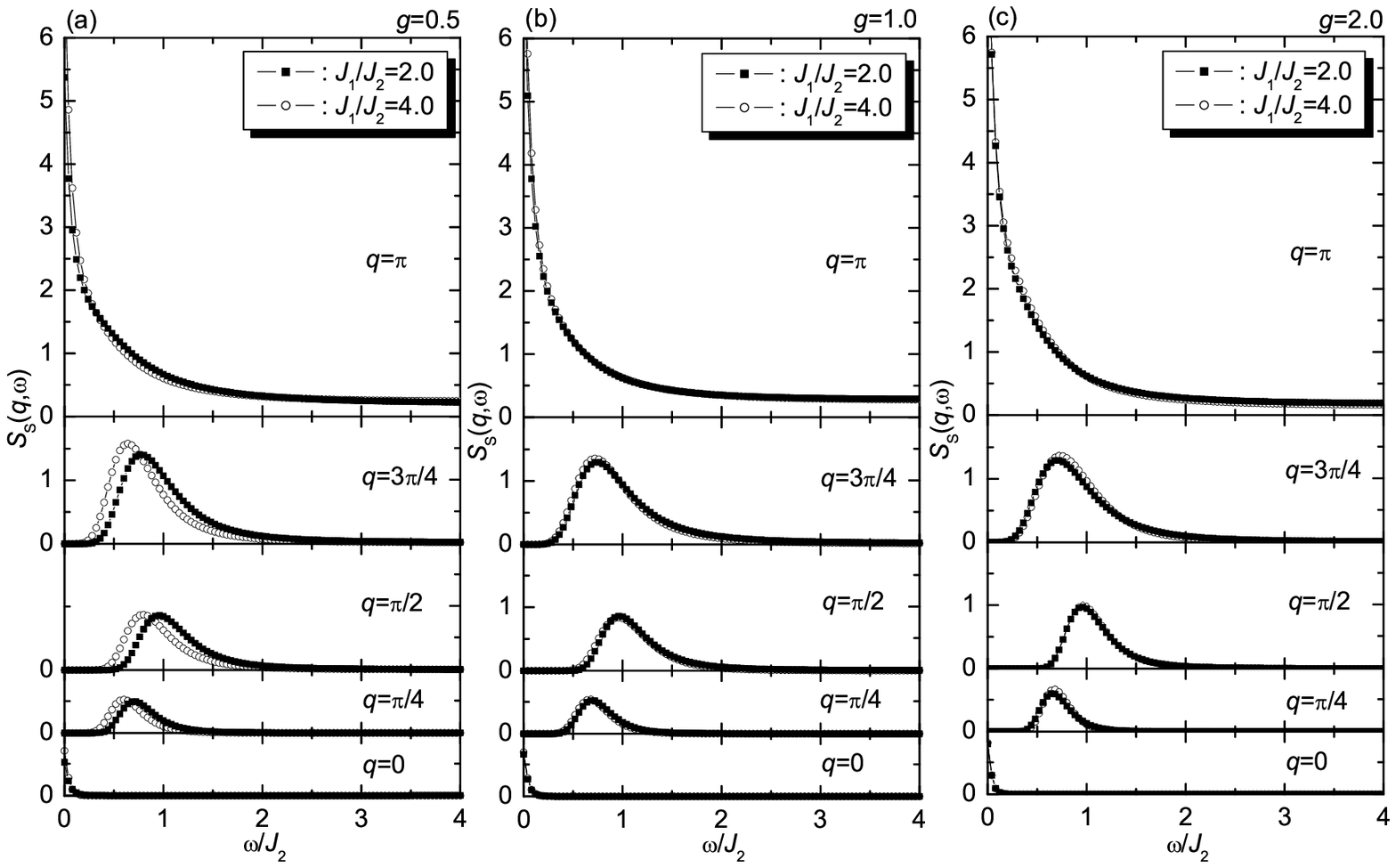}
\caption{Dynamical spin structure factor 
$S_{\rm S}(q,\omega)$ for the coupled 
spin-pseudospin model calculated at 
$k_{\rm B}T=0.1J_2$.}
\end{center}
\label{fig:3}
\end{fullfigure}
%%%%%%%%%%%%%%%
The calculated results for the spin excitation spectra at 
low temperature are shown in Fig.~3, where we find that, 
in contrast to the pseudospin spectra, the spin excitation 
spectra change very little; i.e., the peak position, width, 
as well as the shape of the spectra are not affected by 
the parameter $J_1$ when $g\gtrsim 1$.  When $g$ is small, 
however, the peak position is slightly shifted to lower 
energies with increasing the value of $J_1$ (see Fig.~3 (a)).  

The dispersion relation of the spin and pseudospin excitations 
calculated at low temperature are summarized in Fig.~4, which 
are obtained as the momentum dependence of the peak position 
of the spectra.  
For comparison, we show the dispersion of the quantum 
Ising model in Figs.~4 (a) and (c); the gap opens when $g>1$, 
which is closed at $q=\pi$ when $g\rightarrow 1$, leading 
to the `antiferromagnetic' long-range order 
(or zigzag CO).\cite{nishimoto3,sachdev}  
We note that the gap remains open irrespective of the value 
of $g$ when we include the coupling term $J_2$.  
In the left panels and right panels of Fig.~4, we present 
the same dispersion relations $\omega_q$, but in a different 
energy scales, i.e., $\omega_q/J_1$ and $\omega_q/J_2$.  
We find that, unless $g$ is small, the spin excitation spectra 
are always {\em inside} the charge gap, i.e., inside the gap 
of the pseudospin excitation spectrum.  Thus, when the charge 
gap is large, the energy scale of the spin excitations is 
separated from the high-energy charge excitations.  
With decreasing $g$, however, the energy of the charge 
excitation decreases at the momentum $q=\pi$ to couple with 
the spin excitations.  
%%%% FIG.4 %%%%
\begin{figure}[t]
\vspace{5pt}
\begin{center}
\includegraphics[width=7.0cm,clip]{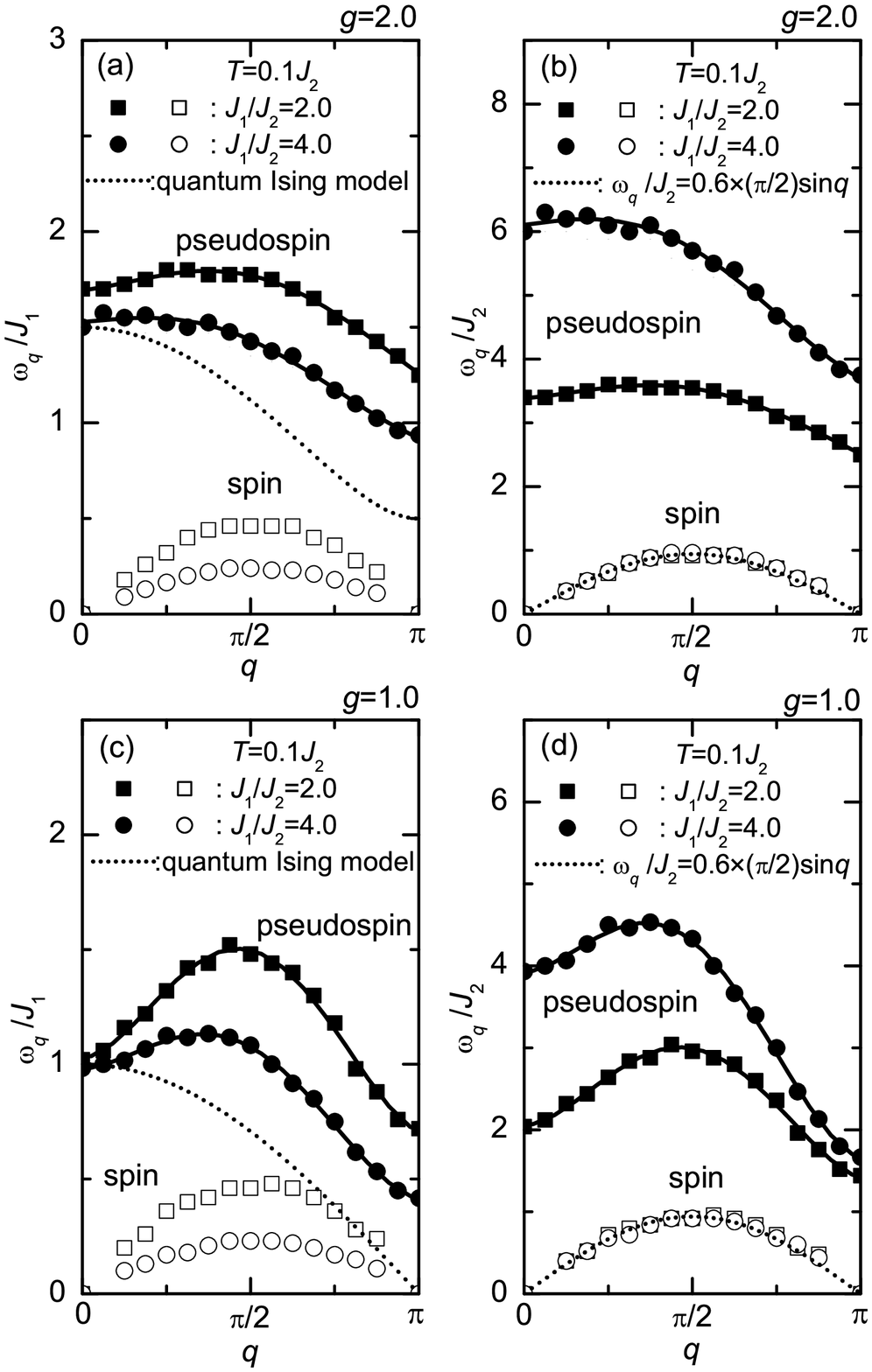}
\caption{Dispersion relations of the spin (open symbols) 
and pseudospin (solid symbols) excitations calculated 
at $k_{\rm B}T=0.1J_2$.  Note that the same data at 
$g=2$ (at $g=1$) are plotted in (a) and (b) (in (c) and 
(d)) in different energy scales $J_1$ and $J_2$.  
The dotted line in (a) and (c) is the dispersion 
relation for the quantum Ising model eq.~(5), and 
that in (b) and (d) is the scaled dispersion 
relation for the 1D antiferromagnetic Heisenberg 
model eq.~(9).}
\end{center}
\label{fig:4}
\end{figure}
%%%%%%%%%%%%%%%

In Fig.~4 (b) and (d), we find that, for $g\gtrsim 1$, the 
dispersion of the spin excitation spectra scales very well 
with $J_2$; i.e., it does not depend on the value of $J_1$.  
More quantitatively, the dispersion of the calculated spin 
excitation spectra is fitted well with the dispersion of 
the 1D antiferromagnetic Heisenberg model 
\begin{equation}
\omega_q/J_2=0.6\times\frac{\pi}{2}\sin q
\end{equation}
if we include the factor $0.6$ as in eq.~(9).  
The factor is independent of $J_1$ for $g\gtrsim 1$ and 
at low $T$.  

These results suggest that at low temperatures there is a 
parameter region where the spin degrees of freedom 
behaves independently from the pseudospin degrees of 
freedom; it is when $g\gtrsim 1$ and the gap of the pseudospin 
excitation spectra is large, inside of which there is 
a spin excitation spectra.  
Thus, we suggest the validity of the decoupling of 
the coupling term of the Hamiltonian as 
\begin{equation}
{\cal H}_{\rm ST}\Rightarrow J_2\sum_{i=1}^L
\big<T_i^+T_{i+1}^-+{\rm H.c.}\big>
\big({\bf S}_i\cdot{\bf S}_{i+1}-\frac{1}{4}\big)
\end{equation}
with
\begin{equation}
\big<T_i^+T_{i+1}^-+{\rm H.c.}\big>\simeq 0.6
\end{equation}
which leads to the effective Heisenberg-model description 
of the spin degrees of freedom of our model.  
Here, it should be noted that in general the factor 
$<T_i^+T_{i+1}^-+{\rm H.c.}>$ at zero temperature takes 
the value $1$ when $g\rightarrow 0$ and $1/2$ when 
$g\rightarrow\infty$; thus the above value $0.6$ reflects 
the effect of quantum fluctuations of pseudospins, which 
is strong already at $g\gtrsim 1$.  It should also be 
noted that with increasing temperature the value of the 
factor decreases to $0$ (due to thermal fluctuations), 
either very rapidly when $g$ is small or rather slowly 
when $g$ is large.  We find that the essential features 
of $<T_i^+T_{i+1}^-+{\rm H.c.}>$ are contained already in 
a two-site (or dimer) model of eq.(1), a minimum model 
reflecting the spin-pseudospin coupling.  
This is evidently shown in Fig.~5.  
%%%% FIG.5 %%%%
\begin{figure}[t]
\vspace{5pt}
\begin{center}
\includegraphics[width=8.0cm,clip]{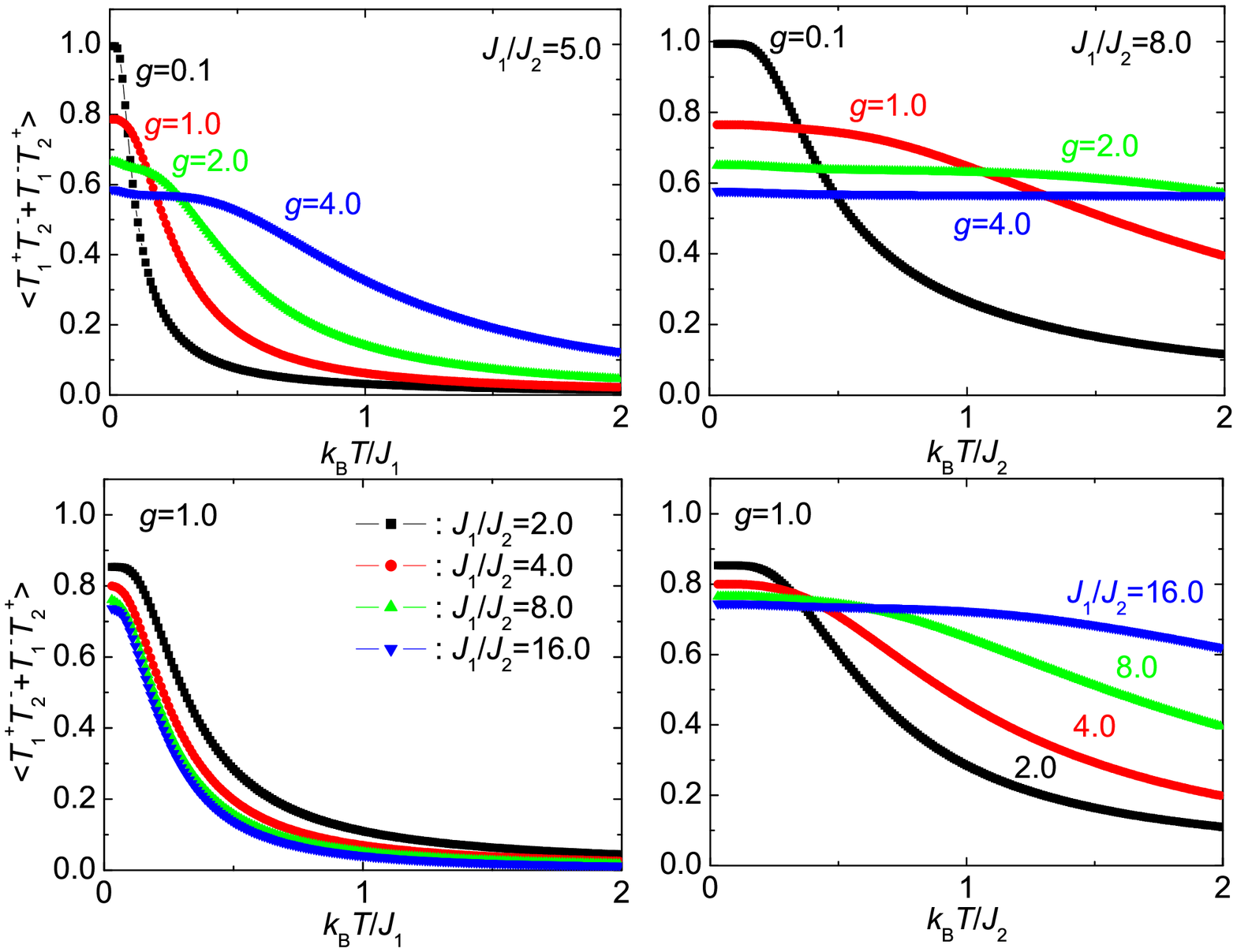}
\caption{Temperature dependence of the pseudospin quantum 
fluctuation $<T_i^+T_{i+1}^-+{\rm H.c.}>$ calculated 
exactly for a dimer model of our spin-pseudospin 
Hamiltonian eq.(1). }
\end{center}
\label{fig:5}
\end{figure}
%%%%%%%%%%%%%%%

%%%%%%%%%%%%%%%%%%%%%%
\subsection{Uniform spin susceptibility}
%%%%%%%%%%%%%%%%%%%%%%

To see the validity of the effective Heisenberg-model 
description further, in particular for its temperature 
dependence, we calculate the temperature dependence of 
the uniform spin susceptibility for the coupled 
spin-pseudospin Hamiltonian.  
The results are shown in Fig.~6, where comparisons are 
made with the uniform susceptibility for the system of 
free spins and with that for the 1D antiferromagnetic 
Heisenberg model.  
We find that the temperature $k_{\rm B}T/J_2$ at which 
$J_2\chi_{\rm S}(T)$ shows a maximum is lower than that 
of the 1D antiferromagnetic Heisenberg model; it becomes 
lower with decreasing the value of $g$ or with increasing 
the value of $J_1/J_2$.  In other words, the deviation 
from the Heisenberg model is large when the quantum 
fluctuation of the pseudospins is small, which occurs 
when $g$ is small or $J_1$ is large.  
%%%% FIG.6 %%%%
\begin{figure}[t]
\vspace{5pt}
\begin{center}
\includegraphics[width=8.5cm,clip]{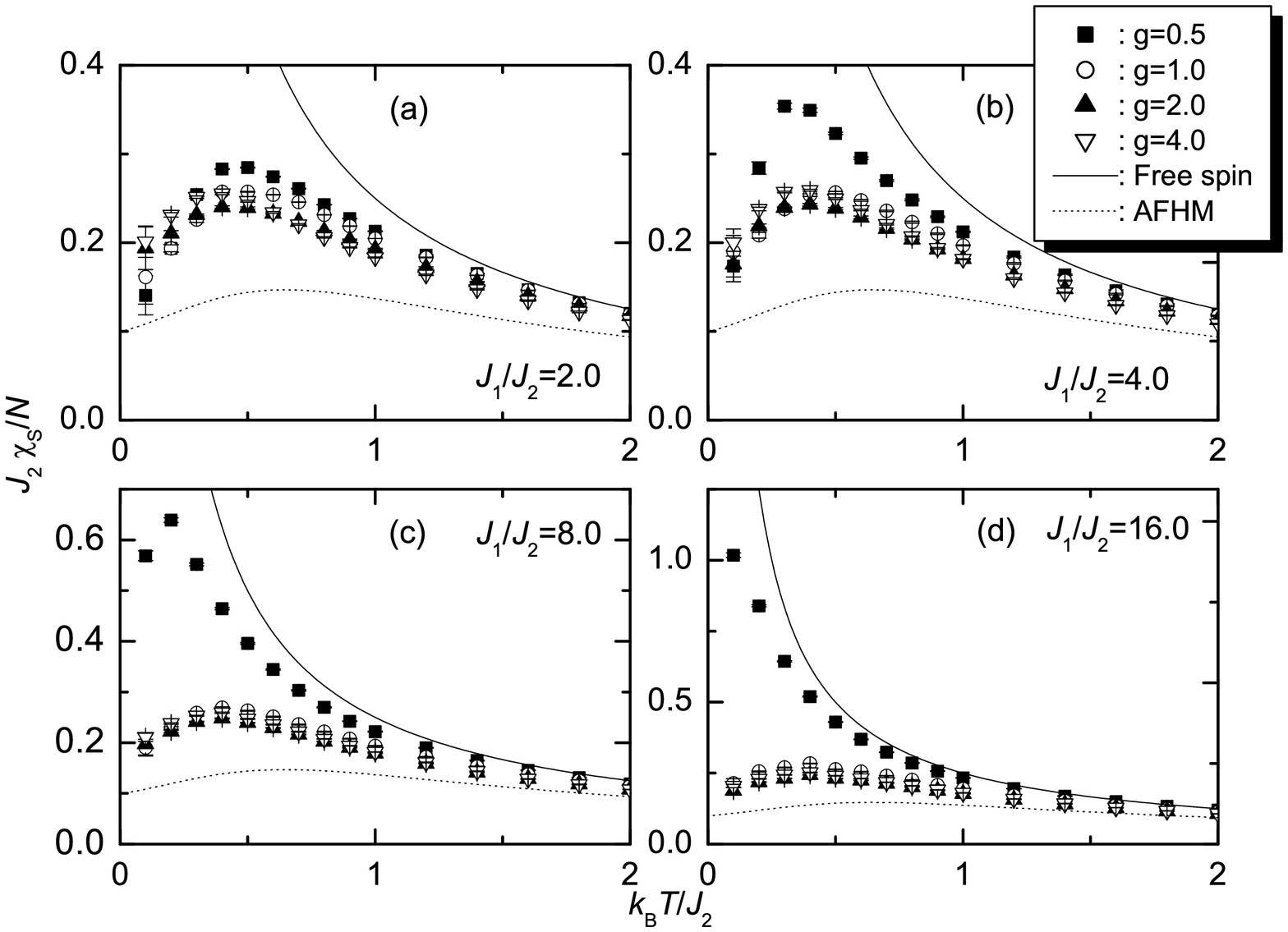}
\caption{Temperature dependence of the uniform spin susceptibility 
$\chi_{\rm S}$ calculated for the coupled spin-pseudospin Hamiltonian.  
The solid and dotted curves are the uniform susceptibility for 
the system of noninteracting $S=1/2$ spins and that for the 1D 
antiferromagnetic Heisenberg model, respectively.}
\end{center}
\label{fig:6}
\end{figure}
%%%%%%%%%%%%%%%

Now, let us analyze the data more precisely.  In order to 
do this, we fit the results with the temperature dependence 
of the spin susceptibility of the 1D antiferromagnetic 
Heisenberg model, the so-called Bonner-Fisher 
curve;\cite{bonner} i.e., we introduce the $T$-dependent 
{\em effective} exchange coupling constant $J_{\rm eff}(T)$ 
and we determine the values so as to fit the calculated 
uniform spin susceptibility $\chi_{\rm S}(T)$.  
If the values of $J_{\rm eff}$ thus obtained do not depend 
on $T$, it follows that the spin degrees freedom of our 
spin-pseudospin model is reduced to a 1D Heisenberg model 
\begin{equation}
{\cal H}_{\rm spin}=J_{\rm eff}\sum_{i=1}^L
{\bf S}_i\cdot{\bf S}_{i+1}
\end{equation}
at least for the response to the uniform magnetic field.  
The results are shown in Fig.~7.  
We find that the estimated value of $J_{\rm eff}(T)$ 
is indeed a constant for temperatures below 
$k_{\rm B}T\lesssim 0.7J_2$ at $g=2$.  A crossover 
temperature $T^*$ $(=0.7J_2)$ is thereby defined.  
The effective exchange coupling constant thus deduced 
takes a value 
\begin{equation}
J_{\rm eff}\simeq 0.6J_2
\end{equation}
at $T<T^*$; this value is consistent with the value 
estimated from the dispersion relation of the spin excitation 
spectra (see \S3.2).  
We find that also at $g=4$ the scaling behavior holds up to 
a higher temperature ($k_{\rm B}T\lesssim 0.8J_2$), but with 
a slightly smaller value of $J_{\rm eff}$ (see Fig.~7 (d)), 
demonstrating the validity of the effective Heisenberg-model 
description at $T<T^*$.  At $g=1$, however, the temperature 
region where $J_{\rm eff}(T)$ takes a constant value is 
already very small, although the value is still 
$J_{\rm eff}\sim 0.6J_2$ at $T\sim 0$ K, and at $g=0.5$, 
the value of $J_{\rm eff}$ at $T\sim 0$ K deviates largely 
from $J_{\rm eff}=0.6J_2$ (or decreases strongly when 
$J_1/J_2$ is large), where the effective Heisenberg-model 
description completely fails.  We thus find that $T^*$ 
thus deduced is insensitive to $J_1$, scales well with 
$J_2$, and depends strongly on $g$ (i.e., $T^*\rightarrow\infty$ 
at $g\rightarrow\infty$ and $T^*\sim 0$ at $g\sim 1$).  
Note that if we assume eq.~(10) the behavior should come 
from the pseudospin fluctuation 
$\left< T^+_iT^-_{i+a}+{\rm H.c.}\right>$, and actually 
we find that very rough tendency in the parameter and 
temperature dependence is seen in the results for the 
dimer model (see Fig.~5) although the scaling behavior 
and the presence of $T^*$ are not seen.  
%%%% FIG.7 %%%%
\begin{figure}[t]
\vspace{5pt}
\begin{center}
\includegraphics[width=8.5cm,clip]{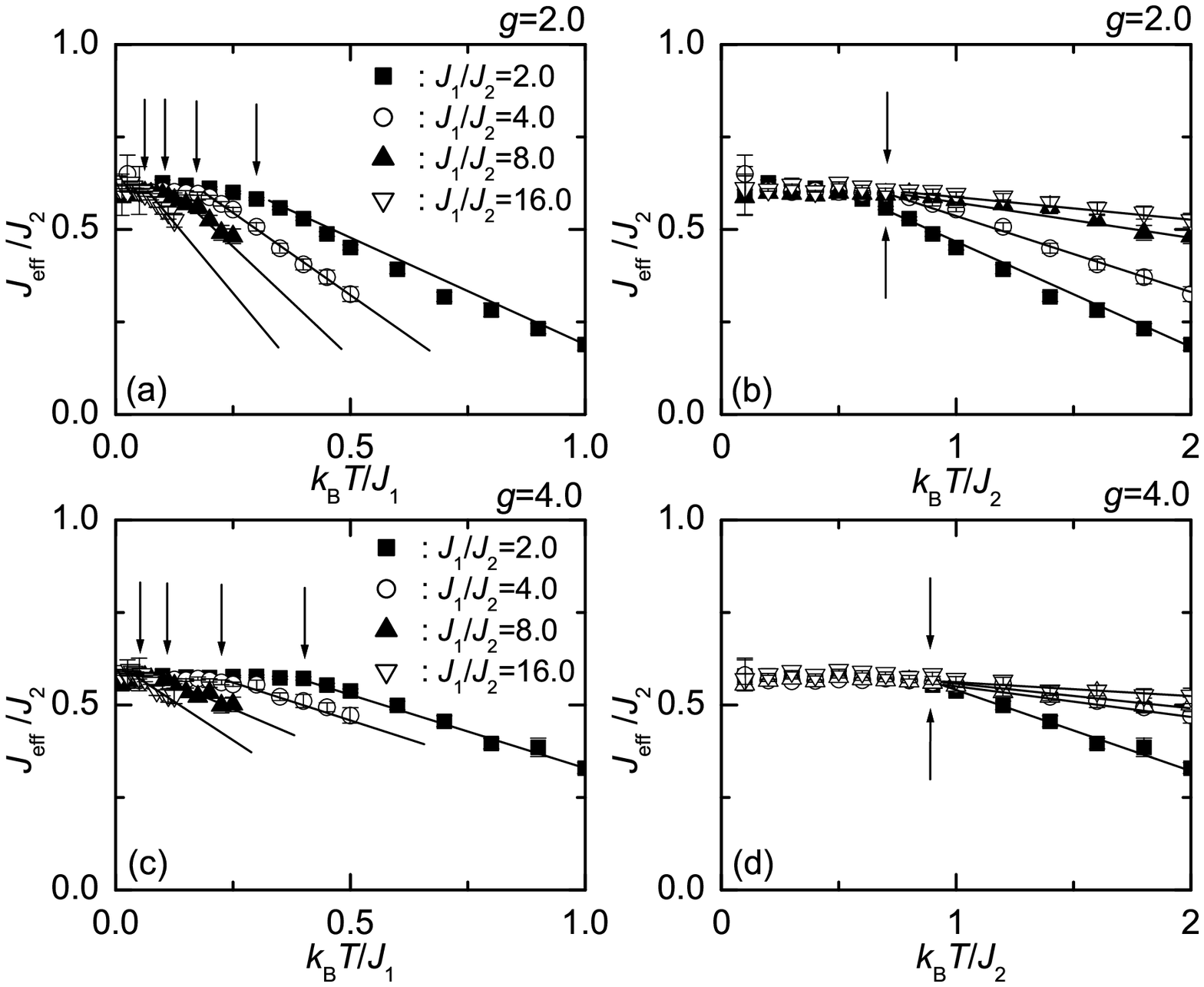}
\caption{Effective exchange coupling constant $J_{\rm eff}(T)$ 
estimated from the fitting of the calculated uniform spin 
susceptibility to the Bonner-Fisher curve.\cite{bonner}  
Note that the same data are plotted as a function of 
$k_{\rm B}T/J_1$ (left panels) and of $k_{\rm B}T/J_2$ 
(right panels), whereby a scaling behavior is seen in the 
latter.  The arrows indicate the crossover temperature $T^*$.  
The solid lines are the guide to the eye.}
\end{center}
\label{fig:7}
\end{figure}
%%%%%%%%%%%%%%%

We note here that the crossover temperature $T^*$ roughly 
scales with $J_2$ rather than $J_1$, as seen in Fig.~7.  
One might suppose that it should scale with the size of 
the charge gap: i.e., up to temperatures corresponding 
to the energy of the lowest charge excitations, with which 
the pseudospins can excite, the spin excitations may be 
written in terms of the 1D antiferromagnetic Heisenberg 
model.  However, as we have discussed in \S3.2, the size 
of the charge gap (if it is defined as a low-energy edge 
of the peak) shows a rather complicated behavior and 
does not simply scale with either $J_2$ or $J_1$.  The 
naive picture thus does not hold.  It may be said however 
that, since there is no other excitations available, the 
deviation from the 1D Heisenberg-model description is 
necessarily due to the pseudospin excitations.  

%%%%%%%%%%%%%%%%%%%%%%
\subsection{Specific heat}
%%%%%%%%%%%%%%%%%%%%%%

Finally, we present the calculated results for the 
temperature dependence of the specific heat $C$ by the 
finite-temperature DMRG method.\cite{dmrg1,dmrg2}  
The results are shown in Fig.~8, where we find that 
the curves have a two-peak structure: e.g., at $g=2$ 
and $J_1/J_2=4$, a rather sharp peak appears at a 
low-temperature region $k_{\rm B}T\simeq 0.3J_2$ 
(which scales with $J_2$) and a broad peak structure 
appears at a high-temperature region 
$k_{\rm B}T\gtrsim 0.8J_2$ (which scales with $J_1$), 
each of which corresponds to the spin and pseudospin 
excitations, respectively.  We also find that the 
shape of the low-temperature peak can be fitted very 
well with the temperature dependence of the specific 
heat of the 1D antiferromagnetic Heisenberg model with 
the effective exchange coupling constant 
$J_{\rm eff}\simeq 0.6J_2$ at $g=2$; this value is 
in accord with the value estimated in \S3.2.  
The temperature at which the deviation in the fitting 
occurs is at $k_{\rm B}T/J_2\simeq 0.5-1$ depending 
on the value of $g$, which is also consistent with 
the estmate from the temperature dependence of the 
uniform spin susceptibility.  
The results thus demonstrate the separation of the 
energy scales between spin and pseudospin degrees of 
freedom and the presence of low-energy `magnetic' 
energy scale as has been pointed out in 
ref.\cite{aichhorn}  
%%%% FIG.8 %%%%
\begin{figure}[t]
\vspace{5pt}
\begin{center}
\includegraphics[width=8.7cm,clip]{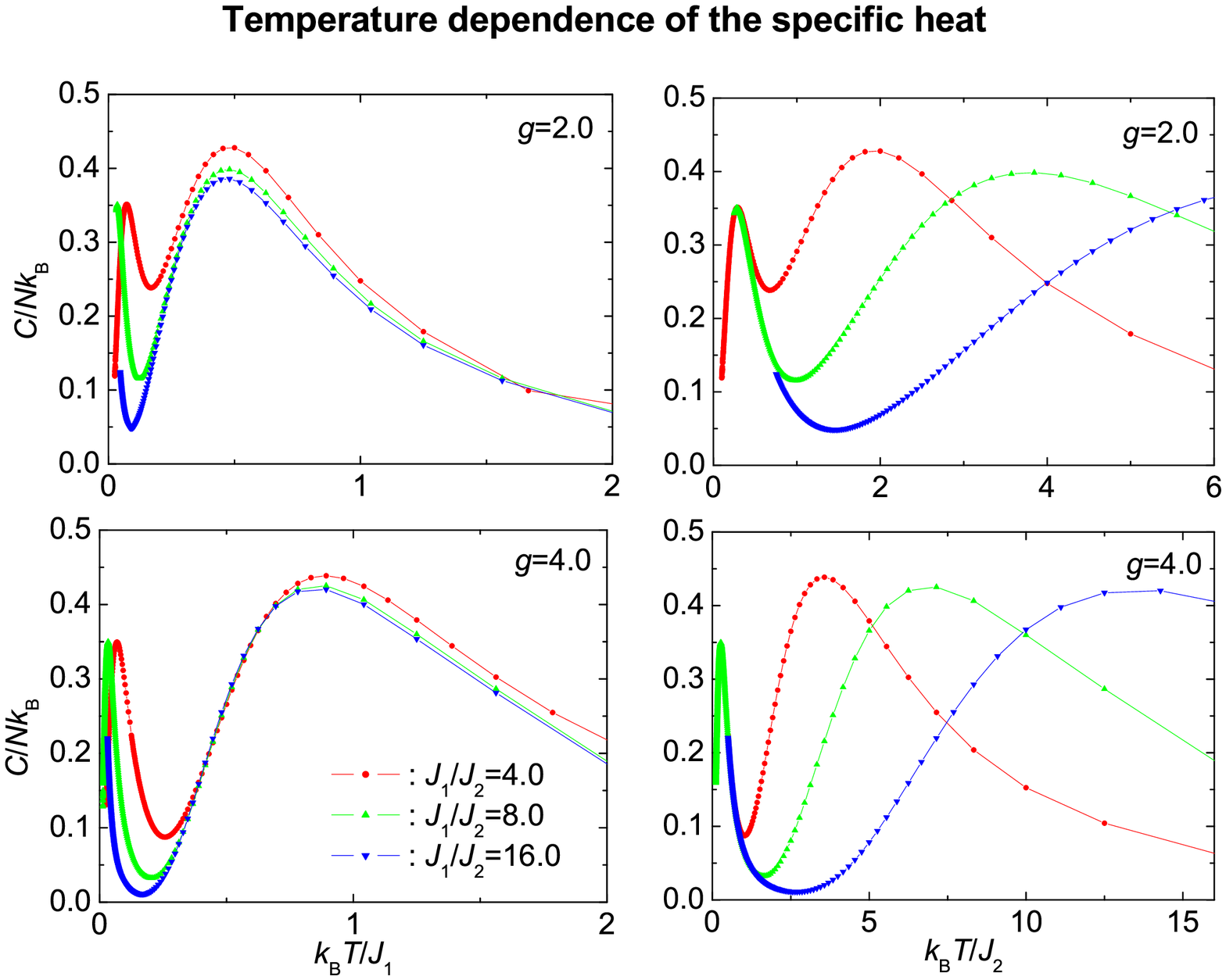}
\caption{Temperature dependence of the specific heat 
$C$ calculated for our coupled spin-pseudospin 
Hamiltonian.}
\end{center}
\label{fig:8}
\end{figure}
%%%%%%%%%%%%%%%

%%%%%%%%%%%%%%%%%%%%%%
\section{Summary and Discussion}
%%%%%%%%%%%%%%%%%%%%%%

We have calculated the spin and pseudospin excitation 
spectra and the temperature dependence of the uniform 
spin susceptibility of the coupled spin-pseudospin 
Hamiltonian by using the QMC method.  We have also 
calculated the temperature dependence of the specific 
heat of the model by the finite-temperature DMRG method.  
We have first shown that, when the pseudospin quantum 
fluctuation is large ($g\gtrsim 1$), the dispersion 
relation of the spin exitation spectra of our model at 
low temperatures agrees well with that of the 1D 
antiferromagnetic Heisenberg model with the renormalized 
effective exchange coupling constant $J_{\rm eff}=0.6J_2$ 
that is independent of the energy scale of the pseudospin 
system $J_1$.  Here, the spin excitation spectra is 
well inside the charge gap, and thus the spin degrees 
of freedom are separated from the charge degrees of 
freedom.  
We then have shown that the temperature dependence of 
the uniform spin susceptibility of our model is well 
described again by the 1D antiferromagnetic Heisenberg 
model with the same effective exchange coupling constant 
$J_{\rm eff}=0.6J_2$.  This description is valid up to 
the crossover temperature $T^*$ that is related to the 
pseudospin excitations of the system and roughly scales 
with $J_2$ unless the quantum fluctuation of the 
pseudospins is small ($g\lesssim 1$).  We have shown 
the appearence of the two-peak structure in the 
temperature dependence of the specific heat and have 
confirmed the presence of the low-energy magnetic 
energy scale.  
We have thus demonstrated the validity of the effective 
Heisenberg-model description of the coupled 
spin-pseudospin model for the quarter-filled ladders.  
It then follows that the coupling between the spin 
and pseudospin degrees of freedom, which occurs at 
$g\lesssim 1$, leads to the {\em anomalous} spin and 
charge dynamics of the system.  

Finally, let us discuss some possible experimental 
relevance of our results.  We have rough estimates 
of the values of the physical parameters appropriate 
for $\alpha'$-NaV$_2$O$_5$; 
according to ref.~\cite{nishimoto1}, we have 
$t_\parallel\sim 0.14$ eV, 
$t_\perp\sim 0.30$ eV, 
and $V_\parallel\sim V_\perp\sim 0.8$ eV, 
which lead to 
$J_1\sim 1.6$ eV, 
$J_2\sim 0.10$ eV, and 
$g\sim 0.75$.  
We thus find that the real material may be in the 
region of $g\lesssim 1$, where the spin degrees of 
freedom are not completely separated from the charge 
degrees of freedom.  The anomalous response of the 
spin degrees of freedom may therefore be expected.  
We here point out, e.g., that the value of 
$J_{\rm eff}$ estimated from the uniform 
susceptibility observed in experiment (which 
takes the value $\sim 600-700$ K at $T\sim 0$ K) 
indeed decreases with increasing temperature,\cite{johnston} 
which is consistent with the results of our 
calculation.  Our explanation is that the effective 
spin exchange coupling constant depends primarily on 
the fluctuation of pseudospins, the temperature 
dependence of which is strong above the crossover 
temperature ($T^*\sim 0$ K in this material) 
where the spin-charge coupling becomes relevant.  
The reported\cite{ohama3} temperature dependence of 
the nuclear spin-lattice relaxation rate $1/T_1$ in 
this material is also interesting in the present 
respect.  The measured\cite{presura} strong temperature 
dependence of the integrated optical spectral weight 
have also been discussed in terms of the destruction 
of short-range spin correlations which occurs as 
the temperature is increased.\cite{aichhorn} 
Although the real material $\alpha'$-NaV$_2$O$_5$ 
is modeled well as a 2D trellis-lattice system 
in particular for its charge degrees of freedom, 
we here want to point out that observed anomalous spin 
dynamics in the disorder phase of this typical CO 
material may possibly be a consequence of this 
type of spin-charge coupling.  

Because the anomalous charge dynamics has been noticed 
also in other transition-metal oxides\cite{amasaki} 
and some organic systems\cite{shibata}, we hope that 
the present study will stimulate further researches on 
the intriguing interplay between the spin and charge 
degrees of freedom of strongly correlated electron 
systems with CO instability.  

\section*{Acknowledgements}
We would like to thank A. W. Sandvik, T. Mutou, 
N. Shibata, Y. Shibata, and T. Suzuki for useful 
discussions on the numerical techniques and T. Ohama 
for enlightening discussion on the experimental aspects.  
This work was supported in part by Grants-in-Aid for 
Scientific Research from the Ministry of Education, 
Culture, Sports, Science, and Technology of Japan.  
Computations were carried out at the computer centers 
of the Research Center for Computational Science, 
Okazaki National Research Institutes, and the Institute 
for Solid State Physics, University of Tokyo.


\begin{thebibliography}{99}

%%% NaV2O5: Theory %%%
\bibitem{smolinski} H. Smolinski, C. Gros, W. Weber, 
U. Pechert, G. Roth, M. Weiden, C. Geibel: 
Phys. Rev. Lett. {\bf 80} (1998) 5146.  
\bibitem{seo} H. Seo and H. Fukuyama: 
J. Phys. Soc. Jpn. {\bf 67} (1998) 2602.  
\bibitem{nishimoto1} S. Nishimoto and Y. Ohta: 
J. Phys. Soc. Jpn. {\bf 67} (1998) 2996.  
\bibitem{thalmeier} P. Thalmeier and P. Fulde: 
Europhys. Lett. {\bf 44} (1998) 242.  
\bibitem{mostovoy1} M. V. Mostovoy and D. I. Khomskii: 
Solid State Commun. {\bf 113} (2000) 159. 

%%% NaV2O5: Experiment %%% 
\bibitem{isobe} M. Isobe and Y. Ueda: 
J. Phys. Soc. Jpn. {\bf 65} (1996) 1178.  
\bibitem{ohama1} T. Ohama, H. Yasuoka, M. Isobe, 
and Y. Ueda: Phys. Rev. B {59} (1999) 3299.  
\bibitem{sawa} H. Sawa, E. Ninomiya, T. Ohama, H. Nakao, 
K. Ohwada, Y. Murakami, Y. Fujii, Y. Noda, M. Isobe, 
and Y. Ueda: J. Phys. Soc. Jpn. {\bf 71} (2002) 385.  
\bibitem{johnston} D. C. Johnston, R. K. Kremer, M. Troyer, 
X. Wang, A. Kl\"umper, S. L. Bud'ko, A. F. Panchula, 
and P. C. Canfield: Phys. Rev. B {\bf 61} (2000) 9558.  
\bibitem{ohama2} For a review, see T. Ohama: 
Bussei Kenkyu (Kyoto) {\bf 74} (2000) 391 
[in Japanese].  

%%% NaV2O5: Anomalous charge fluctuation %%%
\bibitem{ravy} S. Ravy, J. Jegoudez, and A. Revcolevschi: 
Phys. Rev. B {\bf 59} (1999) R681.  
\bibitem{nakao} H. Nakao, K. Ohwada, N. Takesue, Y. Fujii, 
M. Isobe, and Y. Ueda: Physica B {\bf 21-243} (1998) 534.  
\bibitem{damascelli} A. Damascelli, C. Presura, 
D. van der Marel, J. Jegoudez, and A. Revcolevschi: 
Phys. Rev. B {\bf 61} (2000) 2535.  
\bibitem{presura} C. Presura, D. van der Marel, 
A. Damascelli, and R. K. Kremer: 
Phys. Rev. B {\bf 61} (2000) 15762.  
\bibitem{nishimoto2} S. Nishimoto and Y. Ohta: 
J. Phys. Soc. Jpn. {\bf 67} (1998) 3679.  
\bibitem{nishimoto3} S. Nishimoto and Y. Ohta: 
J. Phys. Soc. Jpn. {\bf 67} (1998) 4010.  
\bibitem{mostovoy2} M. V. Mostovoy, J. Knoester, 
and D. I. Khomskii: Phys. Rev. B {\bf 65} (2002) 064412.  
\bibitem{hemberger} J. Hemberger, M. Lohmann, M. Nicklas, 
A. Loidl, M. Klemm, G. Obermeier, and S. Horn: 
Europhys. Lett. {\bf 42} (1998) 661.  
\bibitem{sushkov} A. B. Sushkov, J. L. Musfeldt, X. Wei, 
S. A. Crooker, J. Jegoudez, and A. Revcolevschi: 
Phys. Rev. B {\bf 66} (2002) 054439.  
\bibitem{ohama3} T. Ohama, M. Isobe, and Y. Ueda: 
Physica B {\bf 329-333} (2003) 934.  
\bibitem{aichhorn} M. Aichhorn, P. Horsch, W. von der Linden, 
M. Cuoco: Phys. Rev. B {\bf 65} (2002) 201101.  

%%% Perturbation theory %%%
\bibitem{cuoco} M. Cuoco, P. Horsch, and F. Mack: 
Phys. Rev. B {\bf 60} (1999) R8438.  
\bibitem{sa} D. Sa and C. Gros: 
Eur. Phys. J. B {\bf 18} (2000) 421.  

%%% Nakaegawa: Master thesis %%%
\bibitem{nakaegawa} For details, see T. Nakaegawa: 
{\it Master thesis} (Chiba University, 2002).  

%%% finite-temperature DMRG %%%
\bibitem{dmrg1} X. Wang and T. Xiang, {\it Phys. Rev. B} 
{\bf 56} (1997) 5061.
\bibitem{dmrg2} N. Shibata, {\it J. Phys. Soc. Jpn.} 
{\bf 66} (1997) 2221.

%%% Quantum Ising model %%%
\bibitem{sachdev} S. Sachdev: {\it Quantum Phase Transitions} 
(University Press, Cambridge, 1999).  

%%% Uniform susceptibility %%%
\bibitem{bonner} J. C. Bonner and M. E. Fisher: 
Phys. Rev. {\bf 135} (1964) A640.  

%%% Charge ordering in other systems %%%
\bibitem{amasaki} R. Amasaki, Y. Shibata, and Y. Ohta: 
Phys. Rev. B {\bf 66} (2002) 012502.  
\bibitem{shibata} Y. Shibata, S. Nishimoto, and Y. Ohta: 
Phys. Rev. B {\bf 64} (2001) 235107.  

\end{thebibliography}
\end{document}